\documentclass[11pt]{article}
\usepackage{jcappub}
\usepackage{bm}
\usepackage{color}
\usepackage{array}
\usepackage{graphicx}
\usepackage{multirow}
\newcolumntype{P}[1]{>{\centering\arraybackslash}p{#1}}
\newcolumntype{M}[1]{>{\centering\arraybackslash}m{#1}}

\newcommand{\ud}{\text{d}}
\newcommand{\ER}{E_\text{R}}
\newcommand{\mDM}{m_{\chi}}

\newcommand{\eH}{\mathcal{H}}
\newcommand{\eR}{\mathcal{R}}

\newcommand{\een}{\end{subequations}}
\newcommand{\ben}{\begin{subequations}}

\newcommand{\lsim}{\mathrel{\mathop{\kern 0pt \rlap
      {\raise.2ex\hbox{$<$}}}\lower.9ex\hbox{\kern-.190em $ \sim$}}}
\newcommand{\gsim}{\mathrel{\mathop{\kern 0pt
      \rlap{\raise.2ex\hbox{$>$}}}\lower.9ex\hbox{\kern-.190em $\sim$}}}

\newcommand {\vmin}{v_{\rm min}}

\newcommand{\Ed}{E^{\prime}}
\newcommand{\erf}{\mbox{erf}}
\newcommand{\keVee}{{\rm keVee}}
\newcommand{\GeV}{{\rm GeV}}
\newcommand{\vesc}{v_{\rm esc}}
\newcommand{\Nesc}{N_{\rm esc}}
\newcommand{\vobs}{v_E}

\title{Anapole Dark Matter after DAMA/LIBRA-phase2}

\author[a]{Sunghyun Kang}
\author[a]{Stefano Scopel}
\author[a]{Gaurav Tomar}
\author[a]{Jong--Hyun Yoon}
\affiliation[a]{Department of Physics, Sogang University, Seoul 121-742, South Korea}
\emailAdd{francis735@naver.com}
\emailAdd{scopel@sogang.ac.kr}
\emailAdd{tomar@sogang.ac.kr}
\emailAdd{jyoon@sogang.ac.kr}

\author[b]{Paolo Gondolo}
\affiliation[b]{Department of Physics, University of Utah, 115 South 1400
  East \#201, Salt Lake City, Utah 84112-0830}
\emailAdd{paolo.gondolo@utah.edu}

\abstract{ We re--examine the case of anapole dark matter as an
  explanation for the DAMA annual modulation in light of the
  DAMA/LIBRA--phase2 results and improved upper limits from other DM
  searches. If the WIMP velocity distribution is assumed to be a
  Maxwellian, anapole dark matter is unable to provide an explanation
  of the DAMA modulation compatible with the other
  searches. Nevertheless, anapole dark matter provides a better fit to
  the DAMA--phase2 modulation data than an isoscalar spin--independent
  interaction, due to its magnetic coupling with sodium targets. A
  halo-independent analysis shows that explaining the DAMA modulation
  above 2 keVee in terms of anapole dark matter is basically
  impossible in face of the other null results, while the
  DAMA/LIBRA--phase2 modulation measurements below 2 keVee are
  marginally allowed. We conclude that in light of current
  measurements, anapole dark matter does not seem to be a viable
  explanation for the totality of the DAMA modulation.}

\begin{document}

\maketitle

\section{Introduction}
\label{sec:introduction}

Weakly Interacting Massive Particles (WIMPs) provide one of the most
popular explanations for the Dark Matter (DM) that is believed to make
up 27\% of the total mass density of the Universe~\cite{planck} and
more than 90\% of the halo of our Galaxy. The scattering rate of DM
WIMPs in a terrestrial detector is expected to present a modulation
with a period of one year due to the Earth revolution around the
Sun~\cite{Drukier:1986tm}.  For more than 15 years, the DAMA
collaboration~\cite{dama_2008,dama_2010,dama_2013} has been measuring
a yearly modulation effect in their sodium iodide target. The DAMA
annual modulation is consistent with what is expected from DM WIMPs,
and has a statistical significance of more than $9\sigma$. However, in
the most popular WIMP scenarios used to explain the DAMA signal as due
to DM WIMPs, the DAMA modulation appears incompatible with the results
from many other DM experiments that have failed to observe any signal
so far.

This has prompted the need to extend the class of WIMP models. In
particular, one of the few phenomenological scenarios that have been
shown \cite{anapole_2014} to explain the DAMA effect in agreement with
the constraints from other experiments is Anapole Dark Matter (ADM)
\cite{anapole1,anapole2,anapole3,anapole4}, for WIMP masses
$m_{\chi}\lsim$ 10 \GeV/c$^2$.

Recently the DAMA collaboration has released first results from the
upgraded DAMA/ LIBRA-phase2 experiment \cite{dama_2018}, increasing
the significance of the effect to 12 $\sigma$. The two most important
improvements compared to the previous phases are that now the exposure
has almost doubled and the energy threshold has been lowered from 2
keV electron--equivalent (keVee) to 1 keVee. While for $m_{\chi}\lsim
10$~\GeV/c$^2$ the DAMA phase--1 data where only sensitive to
WIMP--sodium scattering events, the new data below 2 keVee are in
principle also sensitive to WIMP-iodine scattering, for WIMP speeds
below the escape velocity in our Galaxy. This feature has worsened the
goodness of fit of the DAMA data using a standard Spin-Independent
interaction (SI) \cite{freese_2018,dama_2018_sogang}.

In light of the DAMA/LIBRA--phase2 result, in the present paper we
re--examine the ADM scenario. Moreover, compared to the
analyses in \cite{anapole_2014}, we upgrade the constraints from other
direct detection experiments. In this analysis we use results from 
CDEX~\cite{cdex}, 
CDMSlite~\cite{cdmslite_2017}, 
COUPP~\cite{coupp}, 
CRESST-II~\cite{cresst_II,cresst_II_ancillary}, 
DAMIC~\cite{damic}, 
DarkSide--50 ~\cite{ds50}, 
KIMS~\cite{kims_2014}, 
PANDAX-II~\cite{panda_2017}, 
PICASSO~\cite{picasso}, 
PICO-60~\cite{pico60_2015,pico60},  
SuperCDMS~\cite{super_cdms_2017}
and XENON1T~\cite{xenon_1t,xenon_2018}.

The paper is organized as follows. In Section \ref{sec:model} we
summarize the main features of the ADM scenario, providing the
formulas for WIMP direct detection expected rates; our main results
are in Section \ref{sec:analysis}, where we provide an updated
assessment of ADM in light of the DAMA--phase2 data and of the latest
constraints from other direct detection experiments, both assuming a
Maxwellian WIMP velocity distribution and in a halo--independent
approach. Section \ref{sec:conclusions} is devoted to our
conclusions. In Appendix \ref{app:exp} we provide some details on how
the experimental constraints on ADM have been obtained.

\section{The model}
\label{sec:model}

Anapole dark matter (ADM) is a spin--1/2 Majorana particle that interacts with ordinary matter through the
exchange of a standard photon. The ADM--photon interaction Lagrangian density is
\begin{equation}
{\cal L} =\frac{1}{2}\frac{g}{\Lambda^2}\, \bar{\chi}\gamma^{\mu}\gamma^5\chi \, \partial^{\nu}F_{\mu\nu},
  \label{eq:adm_lagrangian}
\end{equation}
where $\chi$ is the ADM field, $F_{\mu\nu}$ is the electromagnetic field strength tensor,  $g$ is a dimensionless coupling constant, and $\Lambda$ is a new physics scale. 

In the nonrelativistic limit, the Hamiltonian for an ADM particle in an electromagnetic field reduces to a coupling between the WIMP spin operator $\vec{S}_{\chi}$ and the curl of
the magnetic field $\vec{B}$, which by Maxwell's equations is proportional to the electromagnetic current density,
$H=-(2g/\Lambda^2) \, \vec{S}_\chi \cdot \vec{\nabla}\times \vec{B}$. 

The nonrelativistic scattering of an 
ADM particle with a nucleon can also be described by the contact interaction Hamiltonian
\begin{equation}
  H_{\chi N} = \frac{2e g}{\Lambda^2} \,  \vec{S}^{\perp}_{\chi}\cdot  \bigg (  e_N \vec{v}_{\chi N}^{\perp} +i \frac{g_N}{2 m_N} \, \vec{q}\cdot
    \vec{S}_N \bigg ) .
  \label{eq:nr_interaction}
\end{equation}
Here $e$ is the elementary charge, $e_N$ is the nucleon charge in units of $e$ ($e_{\rm p}=1$, $e_{\rm n}=0$), $g_N$ is the nucleon
magnetic moment in units of nuclear magnetons $e/2m_N$ ($g_{\rm p}=5.585\,694\,713(46)$, $g_{\rm n}=-3.826\,085\,45(90)$), $m_N$ is the nucleon mass, $\vec{S}_{\chi}$ and
$\vec{S}_{N}$ are the spins of the WIMP and the nucleon, respectively, $\vec{q}$ is the
 momentum transfer, and $\vec{v}^{\perp}_{\chi N}$ is the component of the
WIMP--nucleon relative velocity perpendicular to $\vec{q}$. 

The 
differential cross section per unit nucleus recoil energy $E_R$ for the scattering of an ADM particle of speed $v$ off a nucleus $T$ of mass $m_T$ at rest is given by
\cite{anapole_2014}
\begin{equation}
\frac{\ud\sigma_T}{\ud\ER} = \sigma_{\rm ref}
\frac{m_T}{m_{\chi N}^2} \frac{\vmin^2}{v^2} \left[ Z_T^2 \left(
  \frac{v^2}{\vmin^2} - 1 \right) F_{{\rm E}, T}^2(q^2) + 2
  \mu_T^2 \frac{m_{\chi T}^2}{m_N^2} \left(
  \frac{J_T + 1}{3 J_T} \right) F_{{\rm M}, T}^2(q^2) \right] .
\label{eq:diff_cross_section}
\end{equation}
Here $\vmin$ is the minimum WIMP speed necessary to transfer energy $E_R$, $q^2 = 2
m_T \ER$ is the square of the momentum transfer, $m_{\chi N}$ and $m_{\chi T}$ are the
DM--nucleon and DM--nucleus reduced masses, respectively, $Z_T$ is the atomic number of the nucleus, $\mu_T$ is the magnetic
moment of the nucleus in units of the nuclear magneton $e /
2 m_N$, $J_T$ is the nucleus spin, and we have defined a reference cross
section
\begin{equation}
\label{eq:ADMsigref}
\sigma_{\rm ref} \equiv \frac{2 m_{\chi N}^2 \, \alpha g^2 }{ \Lambda^4} \, ,
\end{equation}
where $\alpha = e^2 / 4 \pi \simeq 1/137$ is the fine structure constant. In an elastic collision $v_{\rm min}$ is given by
\begin{equation}
\label{eq:vmin}
\vmin = \frac{| \vec{q} |}{2 m_{\chi T}} = \sqrt{\frac{m_T \ER}{2 m_{\chi T}^2}} \ .
\end{equation}

In Eq. (\ref{eq:diff_cross_section}), the first term
corresponds to a WIMP interaction with the nuclear charge,
proportional to the electromagnetic longitudinal form factor
$F_{{\rm E},T}^2(q^2)$. The second term corresponds to a WIMP interaction with the nuclear magnetic field,
described by the transverse electromagnetic form factor
$F^2_{{\rm M},T}(q^2)$. Both form factors are normalized to 1 for a
vanishing momentum transfer, i.e., $F^2_{{\rm E},T}(0)=F^2_{{\rm M},T}(0)=1$. 

At the small $q^2$ relevant for our analysis, the charge distribution
gives the dominant contribution to the electric form factor $F_{\rm
  E}^2(q^2)$. Thus $F_{\rm E}^2(q^2)$ is well described by the Helm form factor
\cite{helm}. 

On the other hand, for
the light WIMPs we consider, the magnetic term in
Eq.~(\ref{eq:diff_cross_section}) is negligible for all the nuclei we
include in our analysis with the exception of sodium.
Ref.~\cite{anapole_2014} took the magnetic form factor for sodium from Fig.~31 of Ref.~\cite{donnelly}, 
which is fitted well by the approximate functional form
\begin{equation}
  F_{\rm  M,Na}^2(q^2)=(1-1.15845q^2+0.903442 q^4)\exp(-2.30722q^2)\, ,
  \label{eq:fm_donnelly}
\end{equation}
where $q$ is in units of fm$^{-1}$. As an alternative, we have
considered taking the sodium magnetic form factor from the nuclear
structure functions in Refs.~\cite{haxton1,haxton2}. There, a
WIMP-nucleon contact Hamiltonian is expressed in terms of a set of
Galilean--invariant interaction operators ${\cal O}$, and the nuclear
form factors are expressed in terms of structure functions $W(q^2)$.
The structure functions are available for a selection of nuclei
relevant to DM direct detection, and are obtained from shell--model
calculations. The ADM--nucleon Hamiltonian in
Equation~(\ref{eq:nr_interaction}) can be expressed in the notation of
Refs.~\cite{haxton1,haxton2} as
\begin{equation}
H_{\chi N} = \sum_{\tau=0,1} ( c_8^\tau {\cal O}_8+ c_9^\tau {\cal O}_9) t^\tau,
\end{equation}
where $\tau$ is a nuclear isospin index, $t^\tau$ is an isospin operator ($t^0 = 1$ and $t^1 = \tau_3$), and
\begin{equation}
c_8^\tau = \frac{2eg}{\Lambda^2} \, e^\tau,
\qquad
c_9^\tau = - \frac{eg}{\Lambda^2} \, g^\tau,
\end{equation}
with $e^0=e^1=1$, $g^0=g_{\rm p}+g_{\rm n}$, $g^1=g_{\rm p}-g_{\rm n}$. The magnetic form factor is expressed in terms of the structure functions $W(q^2)$ in Ref.~\cite{haxton2} as
\begin{equation}
  \mu_T^2 \, F_{{\rm M},T}^2(q^2)=\frac{24\pi J_T}{(J_T+1)(2
    J_T+1)} \,  \sum_{\tau\tau^{\prime}} \bigg[ 
    e^{\tau}e^{\tau^{\prime}} W_{\Delta}^{\tau\tau^{\prime}}(q^2)+
    \frac{g^{\tau}g^{\tau^{\prime}}}{16} W_{\Sigma^{\prime}}^{\tau\tau^{\prime}}(q^2)+
    \frac{e^{\tau}g^{\tau^{\prime}}}{2} W_{\Delta\Sigma^{\prime}}^{\tau\tau^{\prime}}(q^2) 
    \bigg].
    \label{eq:fm_haxton}
\end{equation}
In particular, the nuclear magnetic moment $\mu_T$ is obtained by
setting $q^2=0$ in the previous equation.  We have found that the
nuclear structure functions calculated in \cite{haxton2} lead to a
poor prediction of the sodium magnetic moment, $\mu_{\rm \/NaI} \simeq
0.395$, compared to the measured value $\mu_{\rm \/NaI} \simeq
2.218$. For this reason we did not use the nuclear structure functions
in Ref.~\cite{haxton2}, and instead used the sodium magnetic form
factor in Equation~(\ref{eq:fm_donnelly}), as previously done in
Ref.~\cite{anapole_2014}.

\section{Analysis}
\label{sec:analysis}

In direct DM detection searches, the primary observable is the
number of events counted within an interval or region of ``signal'' values, where the ``signal'' values are expressed in electron-equivalent energies (e.g., $\Ed$ in DAMA) or number of photoelectrons (e.g., cS1 and cS2 in XENON1T) or bubble nucleation energies (e.g., $E_{\rm th}$ in PICO). Using an electron-equivalent energy interval $[\Ed_1,\Ed_2]$ as a proxy for other kinds of ``signal'' regions, the expected event rate within $[\Ed_1,\Ed_2]$ per unit detector mass for elastic WIMP scattering off nuclei is given by (see for example Ref.~\cite{del_nobile_generalized} for details)
\begin{equation}
R_{[\Ed_1, \Ed_2]}(t) = \frac{\rho_{\chi}}{\mDM} \sum_T \frac{C_T}{m_T}
\int_0^\infty \ud \ER \, \int_{v_\text{min}(\ER)}^{\infty} \ud v \, f(v, t) \, v \,
\frac{\ud \sigma_T}{\ud \ER}(\ER, v) 
\,\, \epsilon_{[\Ed_1,\Ed_2]}(\ER) .
\label{eq:rate}
\end{equation}
Here $\rho_\chi$ is the DM mass density, $m_\chi$ is the DM particle
mass, $C_T$ is the mass fraction of nuclei $T$ in the target, $f(v,t)$
is the DM speed distribution in the reference frame of the detector,
and $\epsilon_{[\Ed_1,\Ed_2]}(\ER) $ is the total efficiency for
counting nuclear recoil events of energy $\ER$ in the region
$[\Ed_1,\Ed_2]$. The total counting efficiency is generically a
product of the experimental acceptance $\epsilon(\Ed)$ of an event at
``signal'' value $\Ed$, which depends on selection criteria, and the
probability $G_T(\Ed|\ER)$ that a nuclear recoil event of energy $\ER$
produces a ``signal'' $\Ed$,
\begin{equation}
\epsilon_{[\Ed_1,\Ed_2]}(\ER) = \int_{\Ed_1}^{\Ed_2} \ud\Ed \, \epsilon(\Ed) \, G_T(\Ed|\ER) \, .
\end{equation}
The probability density function $G_T(\Ed|\ER)$ depends on the target nucleus, and incorporates the detector resolution function and the mean values $\overline\Ed(\ER)$. The latter can be expressed in terms of quenching factors or scintillation efficiencies (
see Appendix A for details).

By changing the order of integration between $v$ and $\ER$,
Eq.~(\ref{eq:rate}) can be cast into the form
\cite{del_nobile_generalized}
\begin{equation}
\label{eq:hvf}
R_{[\Ed_1, \Ed_2]}(t) =\frac{\rho_{\chi}}{\mDM} \, \sigma_{\rm ref}  \int_0^\infty \ud v \, f(v, t) \, \eH_{[\Ed_1, \Ed_2]}(v) \, ,
\end{equation}
where
\begin{equation}
\label{eq:HT}
\eH_{[\Ed_1, \Ed_2]}(v) =
\sum_T \frac{C_T}{m_T} \int_0^{\ER^{\rm max}(v)} \ud \ER \, \frac{v^2}{\sigma_{\rm ref}} \, \frac{\ud \sigma_T}{\ud \ER} (\ER, v)\, \, \epsilon_{[\Ed_1,\Ed_2]}(\ER) .
\end{equation}
Here $\ER^{\rm max}(v) = 2 m_{\chi T}^2 v^2/m_T$. Defining the velocity integral
\begin{equation}
\eta(\vmin,t)=\int_{\vmin}^\infty \ud v \, \, \frac{f(v, t)}{v} ,
   \label{eq:eta_h}
\end{equation}
%
and integrating Eq. (\ref{eq:hvf}) by parts, the rate can
 be expressed in the form
\begin{align}
\label{eq:R3}
R_{[\Ed_1, \Ed_2]}(t) & = \int_0^\infty \ud\vmin \, \tilde{\eta}(\vmin, t) \,  \eR_{[\Ed_1, \Ed_2]}(\vmin) \, ,
\end{align}
where $\tilde{\eta}(\vmin, t)$ is the rescaled velocity integral
\begin{align}
\tilde{\eta}(\vmin, t) = \frac{\rho_{\chi}}{\mDM} \, \sigma_{\rm ref} \, \eta(\vmin,t),
\label{eq:eta_tilde_h}
\end{align}
and $\eR_{[\Ed_1, \Ed_2]}(\vmin) $ is the response function
\begin{align}
\label{eq:RT}
\eR_{[\Ed_1, \Ed_2]}(\vmin) = \left. \frac{\partial\eH_{[\Ed_1, \Ed_2]}(v)}{\partial v}\right|_{v = \vmin} .
\end{align}
In the following, we simply write $\eR$ when we do not need to specify $[\Ed_1, \Ed_2]$ or $\vmin$.

\begin{figure}
\begin{center}
  \includegraphics[width=0.65\textwidth]{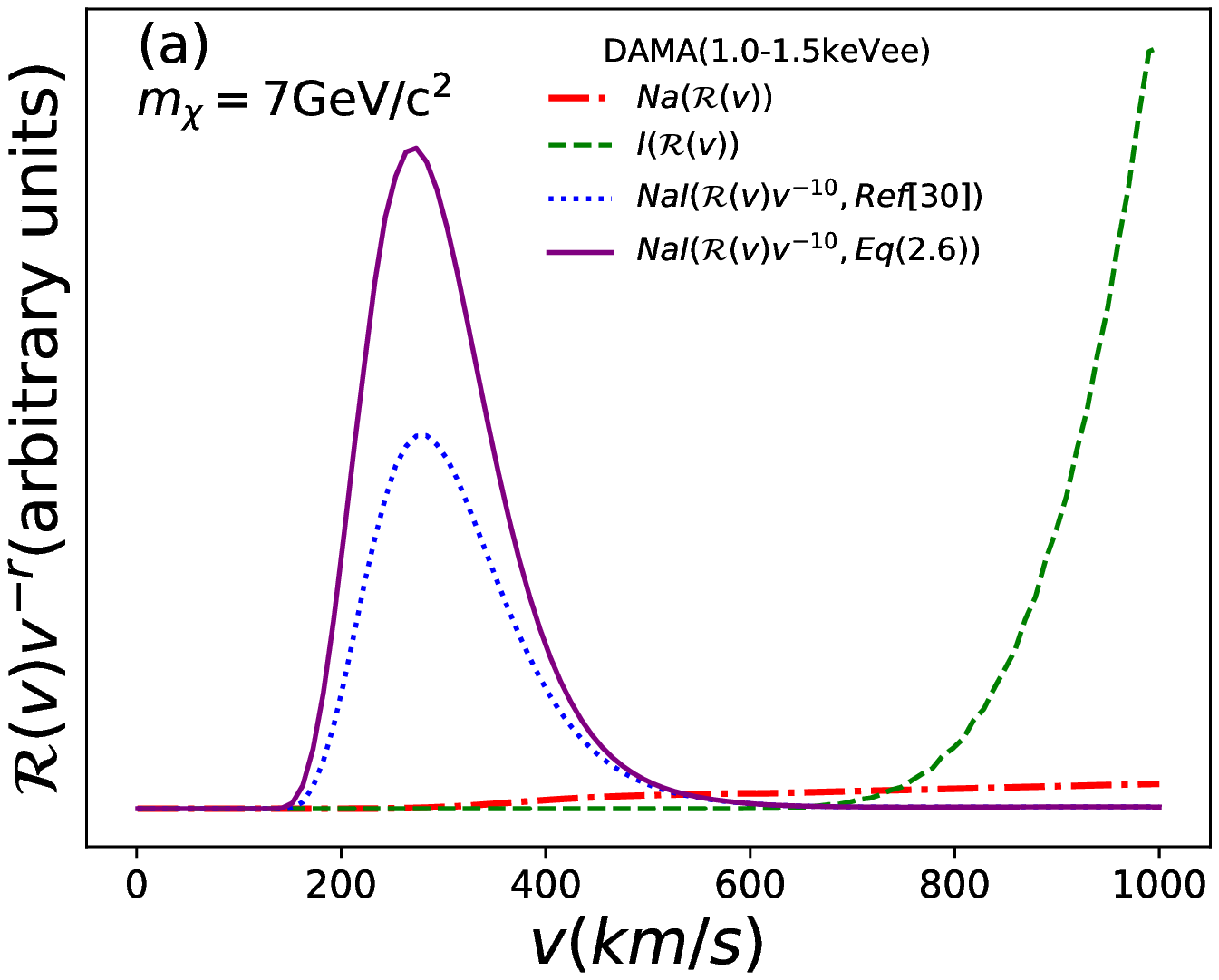}
  \\
  \includegraphics[width=0.65\textwidth]{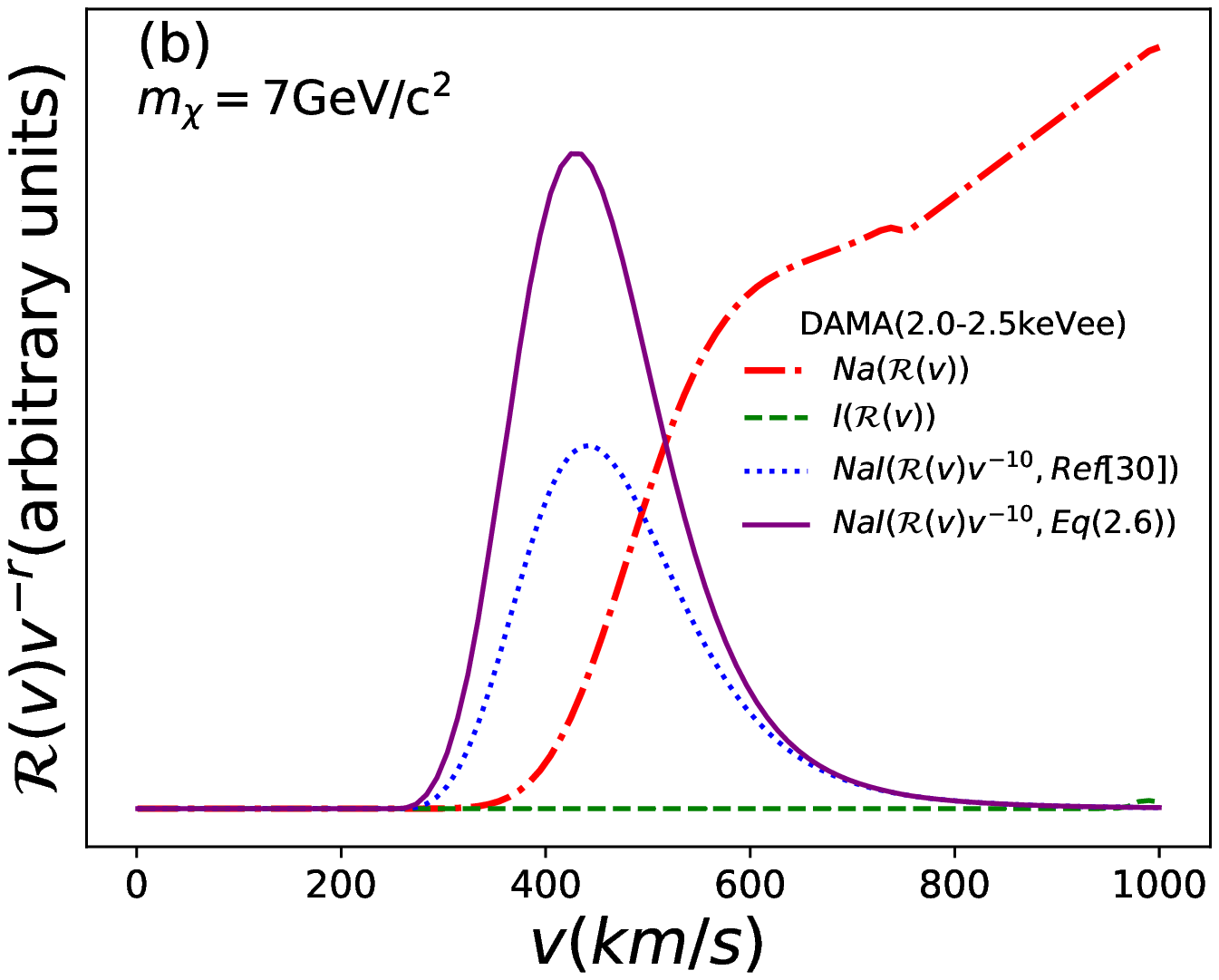}
\end{center}
\caption{DAMA ADM response functions $\eR$ as defined in
  Eq~(\protect\ref{eq:RT}) for $\mDM =7~\GeV/c^2$ in two energy bins:
  (a) $1~\keVee<\Ed<1.5~\keVee$, and (b)
  $2~\keVee<\Ed<2.5~\keVee$. In each plot the dot--dashed lines (red)
  and the dashed lines (green) represent the sodium and the iodine
  contribution, respectively. The solid purple line shows the function
  $v^{-10}\eR$ for NaI used in the regularization procedure of
  Eq.~(\protect\ref{eq:newaverage}), when the magnetic contribution to
  the WIMP--sodium cross section is calculated using
  Eq.~(\protect\ref{eq:fm_donnelly}). The dotted line (blue) shows the
  same quantity but computed using the shell--model calculations of
  the nuclear structure functions in \cite{haxton2},
  Eq.~(\protect\ref{eq:fm_haxton}).\label{fig:r_dama}}
\end{figure}

An explicit example of the response function (\ref{eq:RT}) for DAMA
is provided in Fig.~\ref{fig:r_dama}, where $\eR$ is plotted as a
function of $\vmin$ for $m_{\chi}=7~{\rm \GeV}/c^2$ and for the 
two energy bins (a) $1~\keVee < \Ed < 1.5~\keVee$ and (b) $ 2~\keVee<\Ed <  5.5~\keVee$. 
In both plots, the dot--dashed lines (red)
and the dashed lines (green) represent the contributions to $\eR$
 from WIMP--sodium and WIMP--iodine scattering, respectively.

Due to the revolution of the Earth around the Sun, the velocity
integral $\tilde{\eta}(\vmin, t)$ shows an annual modulation that can
be approximated by the first terms of a harmonic series,
\begin{equation}
\label{etat}
\tilde{\eta}(\vmin, t) = \tilde{\eta}^0(\vmin) +
\tilde{\eta}^1(\vmin) \, \cos\!\left[ \omega (t - t_0) \right] ,
\end{equation}
with $t_0$=2 June being the time of modulation maximum, $\omega = 2
\pi/T$ and $T= 1~{\rm year}$.  As a consequence, the expected rate
shows a similar time dependence
\begin{equation}
\label{eq:Rt}
R_{[\Ed_1, \Ed_2]}(t) = R^0_{[\Ed_1, \Ed_2]} + R^1_{[\Ed_1, \Ed_2]} \, \cos\!\left[ \omega (t - t_0) \right] .
\end{equation}
The two components $\tilde{\eta}^0$ and
$\tilde{\eta}^1$
respectively drive the unmodulated (i.e., time-averaged) part of the
DM signal (to which all direct detection experiments are sensitive)
and the modulated part of the DM signal (the measurement of which requires large exposures and
good detector stability, and represents a possible explanation of the
annual modulation observed by the DAMA experiment).

\subsection{Maxwellian velocity distribution}
\label{sec:maxwellian}

In this Section we assume that the WIMP velocity distribution
in the Galactic rest
frame is a standard isotropic Maxwellian at rest, truncated at the escape velocity $\vesc$,
\begin{equation}
f_{\rm gal}(u) = \frac{1}{\pi^{3/2} v_0^3N_{\rm esc}} \, e^{-u^2/v_0^2} \, \Theta(\vesc - u ).
\end{equation}
Here $u$ is the WIMP speed in the Galactic rest frame, $\Theta$ is the Heaviside step function, and
\begin{equation}
\Nesc = \erf(z)-2 \, z \, e^{-z^2} / \pi^{1/2}
\end{equation}
with $z=\vesc/v_0$.
The WIMP speed distribution in the laboratory frame can be obtained with a change of reference frame. It depends on
the speed of the Earth with respect to the Galactic rest frame, which
neglecting the ellipticity of the Earth orbit, is given by
\begin{align}
\vobs(t) & = \big[ v_{\odot}^2 + v_{\oplus}^2 + 2 \, v_{\odot} \, v_{\oplus} \, \cos\gamma \, \cos[\omega(t-t_0)] \big]^{1/2} 
.
\label{eq:maxwellian}
\end{align}
In this formula,  $v_{\odot}$ is the speed of the Sun in the Galactic rest frame, $v_{\oplus}$ is the speed of the Earth relative to the Sun, and $\gamma$ is the ecliptic latitude of the Sun's motion in the Galaxy. We take $\cos\gamma \simeq 0.49$, 
$v_{\oplus}=2\pi(1~{\rm AU})/(1~{\rm year}) \simeq 29~{\rm km/s}$, 
$v_{\odot}=v_0+12~{\rm km/s}$, $v_0=220~{\rm km/s}$~\cite{v0_koposov}, and
$v_{esc}=550~{\rm km/s}$~\cite{vesc_2014}.

The velocity integral $\eta(\vmin,t)$ for the truncated Maxwellian distribution can finally be computed from the expression of the speed distribution. We have obtained its modulated and unmodulated parts by expanding $\eta(\vmin,t)$ to first order in $v_{\oplus}/v_{\odot}$.

\begin{figure}
\begin{center}
  \includegraphics[width=0.8\textwidth]{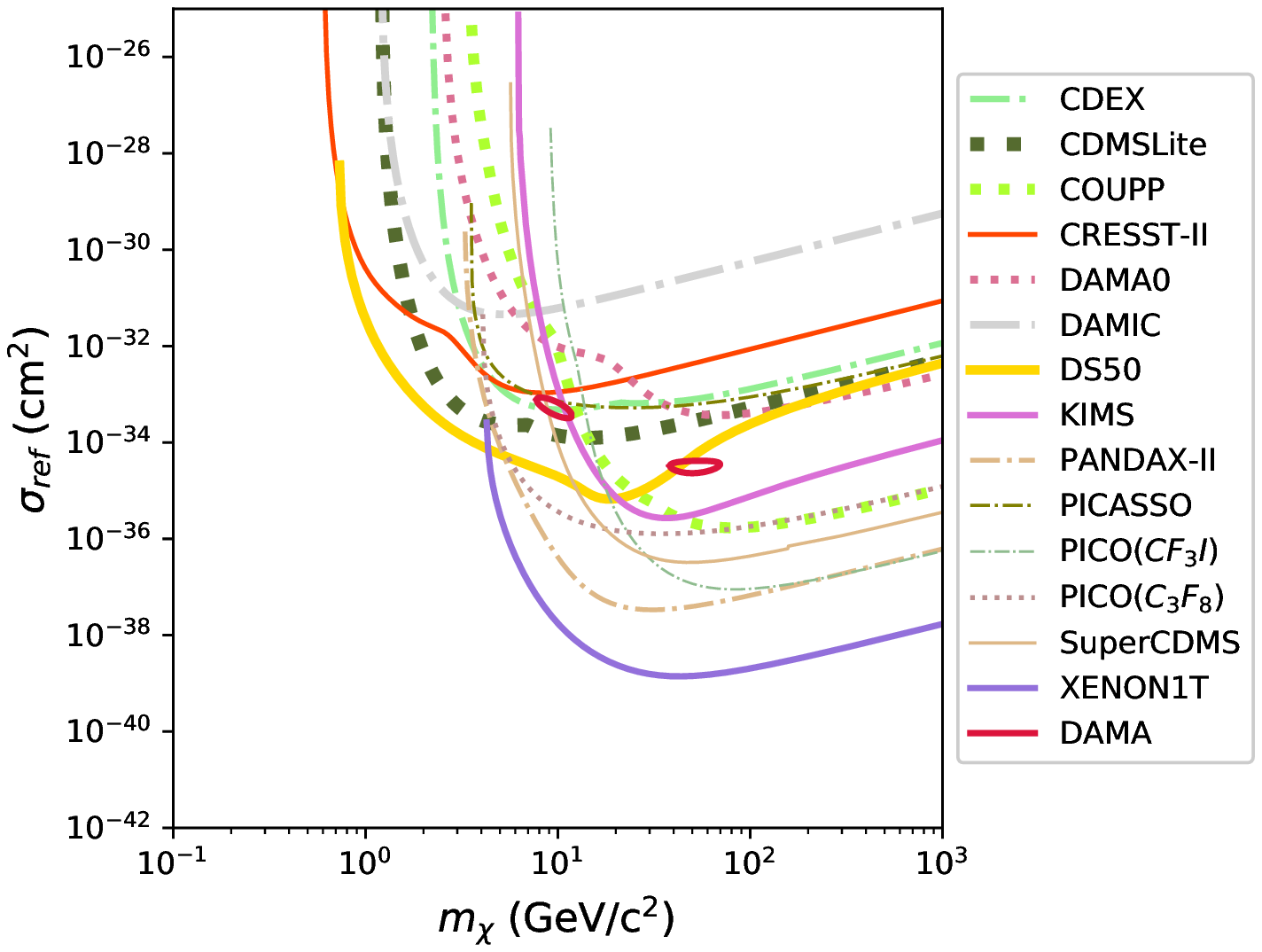}
  \caption{Anapole Dark Matter $5\sigma$ DAMA modulation
    regions in the $\mDM$--$\sigma_{\rm ref}$ plane (inside the two closed solid red
    lines) when the WIMP velocity
    distribution is taken as a standard Maxwellian, and 90\% C.L. upper bounds on
    $\sigma_{\rm ref}$ from other DM direct searches (other
    lines).\label{fig:maxwellian}}
  \end{center}
\end{figure}

\begin{figure}
\begin{center}
  \includegraphics[width=0.7\textwidth]{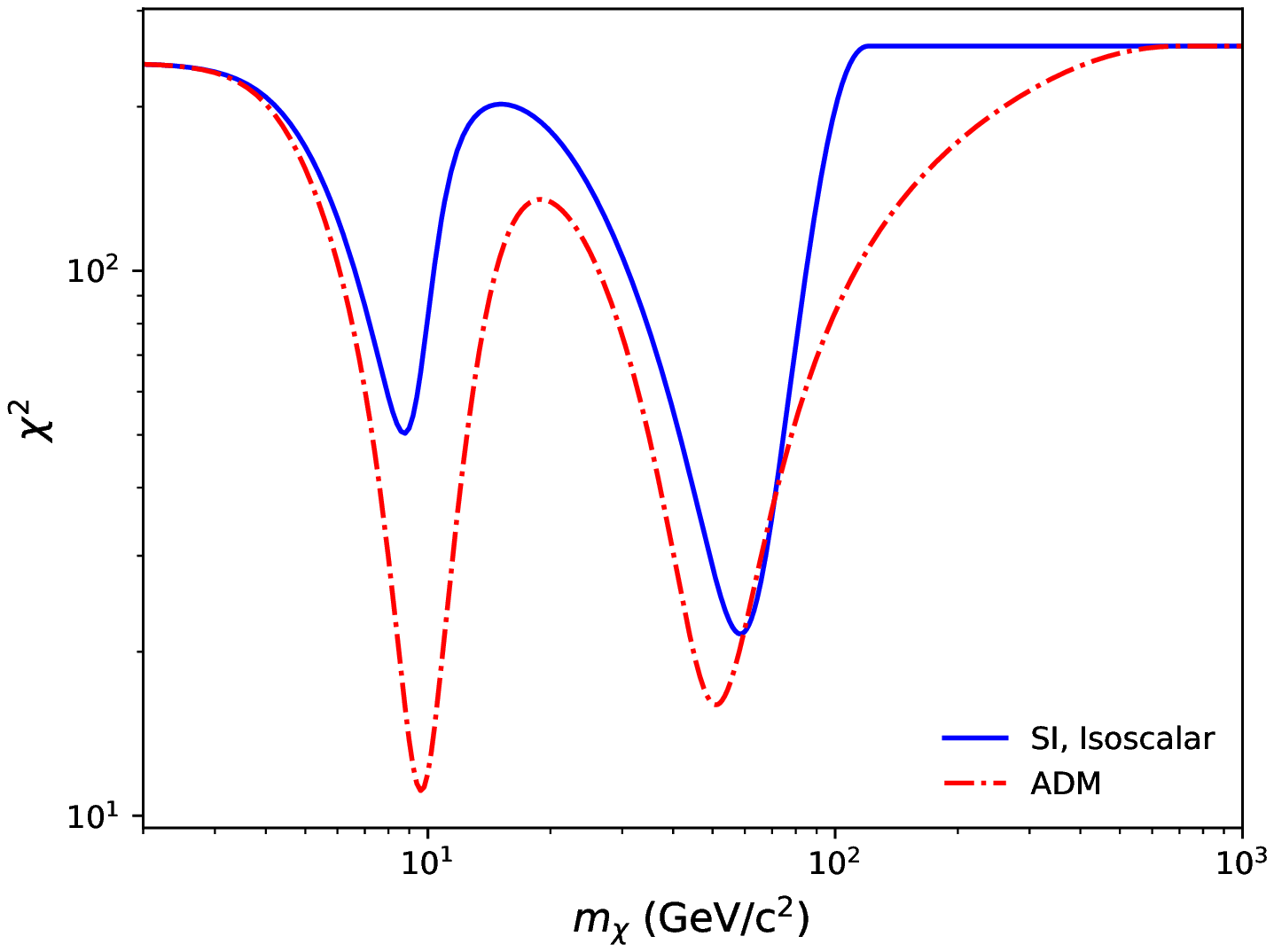}
\end{center}
\caption{Minimum of $\chi^2(\mDM,\sigma_{\rm ref})$ in Eq.~(\ref{eq:chi2}) at fixed
  $\mDM$ for anapole dark matter (red dot--dashed line) and a spin-independent isoscalar
  interaction (blue solid line).\label{fig:adm_mchi_chi2}}
\end{figure}

The expected modulation amplitude $S_{{\rm m},k}$ in the $k$-th DAMA
bin depends on the WIMP mass $m_{\chi}$ and on the reference cross
section $\sigma_{\rm ref}$ (actually, on the product $\sigma_{\rm ref}
\rho_{\chi}$; we use a reference value $\rho_{\chi} =
0.3~\GeV/c^2/{\rm cm}^3$).  To check how well ADM with a Maxwellian
distribution fits the DAMA/LIBRA--phase2 data $S^{\rm exp}_{{\rm m},k}
\pm \sigma_k$ in~\cite{dama_2018}, we perform a $\chi^2$ analysis
constructing the quantity
\begin{equation}
\chi^2(m_{\chi},\sigma_{\rm ref})=\sum_{k=1}^{15} \frac{\left [S_{{\rm m},k}(m_{\chi},\sigma_{\rm ref})-S^{\rm exp}_{{\rm m},k} \right ]^2}{\sigma_k^2}.
  \label{eq:chi2}
  \end{equation}
We consider 14 energy bins of width 0.5 keVee 
from 1 keVee to 8 keVee, and one wider high--energy control bin extending from 8 keVee
to 16 keVee.

The global minimum of $\chi^2(m_{\chi},\sigma_{\rm ref})$ for ADM
occurs at $m_{\chi}=9.6~\GeV/c^2$, $\sigma_{\rm ref} = 5.5 \times
10^{-34}~{\rm cm}^2$, and its value is $\chi^2_{min}=11.1$
($\text{$p$-value}=0.602$ with $15-2$ degrees of freedom, which is an
indication of a good fit).

The $5\sigma$ DAMA modulation regions are plotted in
Fig.~\ref{fig:maxwellian} as the two regions inside the contour
$\chi^2(m_{\chi},\sigma_{\rm ref})=\chi^2_{min}+5^2$ in the
$m_{\chi}$--$\sigma_{\rm ref}$ plane. The region at lower masses
(around $m_\chi \approx 10~\GeV/c^2$ and $\sigma_{\rm ref} \approx
10^{-33}~{\rm cm}^2$) contains the global minimum of the $\chi^2$. The
region at higher masses (around $m_\chi \approx 50~\GeV/c^2$ and
$\sigma_{\rm ref} \approx 3 \times 10^{-35}~{\rm cm}^2$) contains a
secondary local minimum. The other lines in Fig.~\ref{fig:maxwellian}
are the 90\% upper bounds from other existing DM direct--detection
experiments (the region above each line is excluded). As expected, and
in agreement with \cite{anapole_2014}, a DAMA explanation in terms of
ADM is excluded by the null results of other experiments, for a
Maxwellian WIMP velocity distribution.

In the rest of this section we compare the ADM $\chi^2$ to the $\chi^2$ of the often-quoted isoscalar spin-independent case, and we comment on the relative importance of scattering off sodium vs iodine.

In Fig.~\ref{fig:adm_mchi_chi2}, the minimum of
$\chi^2(m_{\chi},\sigma_{\rm ref})$ in Eq.~(\ref{eq:chi2}) at fixed
$m_\chi$ is plotted as a dot--dashed line (red) as a function of
$m_\chi$ (this is the $\chi^2$ obtained by profiling out $\sigma_{\rm
  ref}$). The global minimum around $m_\chi \approx 10~\GeV/c^2$ and
the secondary minimum around $m_\chi \approx 50~\GeV/c^2$ are clearly
visible. Fig.~\ref{fig:adm_mchi_chi2} also shows (solid blue line) the
minimum of the $\chi^2$ at fixed $m_{\chi}$ for an isoscalar
spin-independent (SI) cross section, which scales with the square of
the nuclear mass number. Also in this case two local minima are
present. However now the absolute minimum is the one with the largest
mass~\cite{freese_2018}, while the low--mass local minimum at
$m_{\chi}$=8.8 GeV/c$^2$ has a $\chi^2$ significantly larger
($\chi^2_{min}=50.3$) than in the ADM case.

The different behavior of the $\chi^2$ in the ADM and SI cases can be
understood from the different hierarchy of the WIMP--iodine and
WIMP--sodium cross sections in the two cases. In fact, differently
from the situation with DAMA--phase1, in the two additional
low--energy bins from 1 keVee to 2 keVee of DAMA--phase2 the
modulation effect receives a contribution from WIMP scattering off
iodine targets also at low WIMP masses (below $10~\GeV/c^2$). This can
be seen in Fig.~\ref{fig:r_dama}(a) for $m_{\chi}$=7 GeV/c$^2$, where
for 1 keVee$<E^{ \prime}<$1.5 keVee the contribution to $\eR$ from
WIMP--iodine scattering is different from zero when $\vmin\gsim$ 700
km/s, a range below the escape velocity in the lab frame (for our
standard choice of the astrophysical parameters summarized after
Eq.~(\ref{eq:maxwellian})).  On the other hand, as shown in
Fig.~\ref{fig:r_dama}(b), for the same WIMP mass and for $2~\keVee<E^{
  \prime}<2.5~\keVee$ (i.e., in the lowest energy bin of DAMA--phase1)
WIMP--iodine scattering does not contribute to the expected signal
until the WIMP speed is well above the escape velocity. For SI
interactions, due to the large hierarchy between the WIMP--iodine and
WIMP--sodium cross sections, the additional contribution from
WIMP--iodine scattering is known to lead to a steep rise of the
expected modulation amplitudes for $E^{\prime}< 2~\keVee$. This rise
is incompatible with the DAMA--phase2 measurements, worsening
considerably the goodness--of--fit in going from DAMA--phase1 to
DAMA--phase2 \cite{freese_2018,dama_2018_sogang}.  On the other hand,
in the case of ADM, for low WIMP masses the cross section in
Eq.~(\ref{eq:diff_cross_section}) takes its dominant contribution from
the magnetic component in sodium. This can be seen by examining the
NaI response functions $\eR$ in Fig.~\ref{fig:r_dama}. The response
functions $\eR$ (appropriately multiplied by the factor $v^{-10}$, see
next Section) are the solid purple line when $F^2_{\rm m,Na}(q^2)$ is
evaluated using Eq.(\ref{eq:fm_donnelly}), and the dotted line (blue)
when $F^2_{\rm m,Na}(q^2)$ is evaluated using the nuclear structure
functions of Ref.~\cite{haxton2}.  As pointed out in
Section~\ref{sec:model}, the form factors in \cite{haxton2} largely
underestimate the measured sodium magnetic moment, so that in such
case the WIMP--sodium cross section is only due to the electric
part. Thus, in Fig.~\ref{fig:r_dama}, the difference between the two
evaluations of $\eR$ (solid purple line vs dotted blue line) is due to
the magnetic component in sodium. This implies that for ADM, when the
magnetic contribution of Na is included, it is the dominant one.  In
particular, near the absolute minimum of the $\chi^2$, the enhancement
of the sodium contribution due to the magnetic component of the cross
section reduces the hierarchy between the ADM WIMP--iodine and
WIMP--sodium cross sections compared to the SI case. It is this that
produces a fit of better quality.  The ADM $\chi^2$ in
Fig.~\ref{fig:adm_mchi_chi2} also shows a milder rise at large
$m_{\chi}$ compared to the SI interaction. This is a general property
of models for which the cross section depends explicitly on the WIMP
incoming velocity. This kind of models provides a better fit to the
DAMA modulation data at large values of $m_{\chi}$ than models with a
velocity--independent cross--section, due to the different phase of
the modulation amplitudes~\cite{dama_2018_sogang}.

\subsection{Halo--independent analysis}
\label{sec:halo_independent}

In the halo--independent method of Refs.~\cite{del_nobile_generalized,anapole_2014}, measured rates $R^i_{[\Ed_1, \Ed_2]}$ (with $i=0,1$) are
mapped into 
suitable averages of the two halo functions $\tilde{\eta}^i$. Ref.~\cite{del_nobile_generalized} defines 
averages $\overline{\tilde{\eta}^i}_{[v_{{\rm min},1},
    v_{{\rm min},2}]}$ ($i=0,1$) using $\eR(\vmin)$ in Eq.~(\ref{eq:RT}) as a weight function,
\begin{equation}
\overline{\tilde{\eta}^i}_{[v_{{\rm min},1},v_{{\rm min},2}]} =
\frac{\int_0^\infty \ud\vmin \, \tilde{\eta}^i(\vmin) \, \eR_{[\Ed_1, \Ed_2]}(\vmin)}{\int_0^\infty \ud\vmin \, \eR_{[\Ed_1, \Ed_2]}(\vmin)} .
\label{eq:eta_average}
\end{equation}
The velocity intervals $[v_{{\rm min},1}, v_{{\rm min},2}]$ are
defined as those velocity intervals where the weight function
$\eR_{[\Ed_1, \Ed_2]}(\vmin)$ is sizeably different from zero.

In the case of ADM, the integral in the denominator of
Eq.~(\ref{eq:eta_average}) diverges because the differential cross
section in Eq.~(\ref{eq:diff_cross_section}) depends on a power of $v$
larger than $-2$. Ref.~\cite{anapole_2014} found a solution to this
complication by using the weight functions $\vmin^{-r} \eR(\vmin)$ in
place of $\eR(\vmin)$, where $r$ is a suitable integer. Regularized
averages of $\tilde{\eta}^i$ ($i=0,1$) are defined as
\begin{equation}
\vmin^{-r} \langle \vmin^{r} \tilde{\eta}^i\rangle_{[v_{{\rm min},1},v_{{\rm min},2}]} =
\frac{ \int_0^\infty \ud\vmin \, \vmin^r \, \tilde{\eta}^i(\vmin) \, w^{(r)}_{[\Ed_1, \Ed_2]}(\vmin) }
{\vmin^{r} \int_0^\infty \ud\vmin \, w^{(r)}_{[\Ed_1, \Ed_2]}(\vmin) },
\label{eq:newaverage}
\end{equation}
where
\begin{equation}
w^{(r)}_{[\Ed_1, \Ed_2]}(\vmin) = \vmin^{-r} \, \eR_{[\Ed_1, \Ed_2]}(\vmin) .
\end{equation}
The velocity intervals $[v_{{\rm min},1}, v_{{\rm min},2}]$ are
defined as those velocity intervals where the weight function
$w^{(r)}_{[\Ed_1, \Ed_2]}(\vmin)$ is sizeably different from zero.

Estimates of the regularized averages $ \vmin^{-r} \langle \vmin^{r} \tilde{\eta}^i\rangle_{[v_{{\rm min},1},v_{{\rm min},2}]} $ from measurements $R^i_{[\Ed_1, \Ed_2]}$ are obtained by noticing that the numerator in Eq.~(\ref{eq:newaverage}) is equal to the numerator in Eq.~(\ref{eq:eta_average})  and is equal to $R^i_{[\Ed_1, \Ed_2]}$ by Eqs.~(\ref{eq:R3}) and~(\ref{eq:Rt}),
\begin{equation}
\vmin^{-r} \langle \vmin^{r} \tilde{\eta}^i\rangle_{[v_{{\rm min},1},v_{{\rm min},2}]} =
\frac{ R^i_{[\Ed_1, \Ed_2]} }
{\vmin^{r} \int_0^\infty \ud\vmin \, \vmin^{-r} \, \eR_{[\Ed_1, \Ed_2]}(\vmin) } .
\label{eq:eta_crosses}
\end{equation}

\begin{figure}
\begin{center}
  \includegraphics[width=0.8\textwidth]{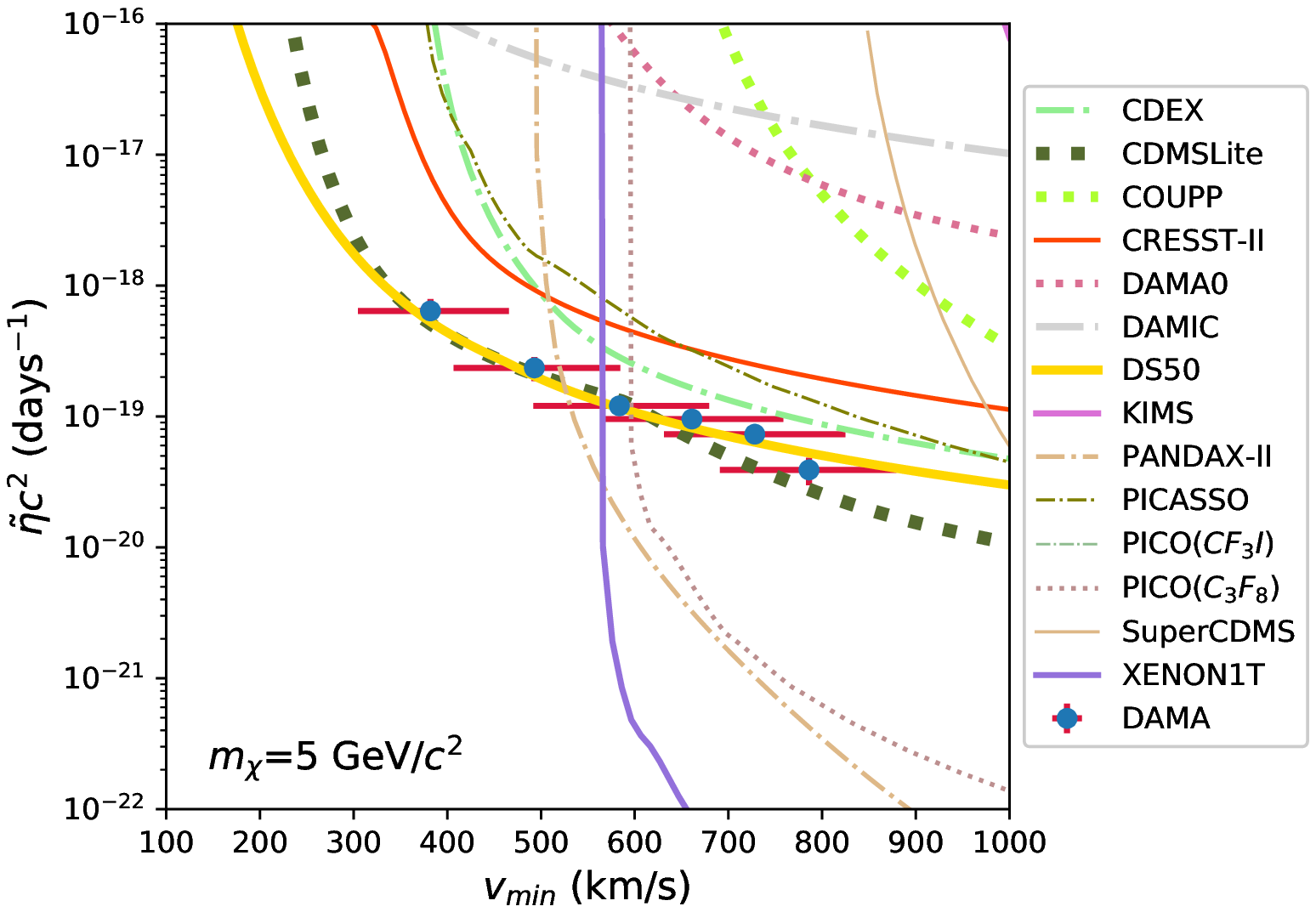}
  \caption{Measurements of $\vmin^{-10} \langle\vmin^{10} \tilde\eta^1(\vmin)\rangle$ (DAMA) and upper bounds on $\tilde\eta^0$ (other experiments)
  for anapole dark matter
  with $\mDM = 5~\GeV/c^2$.\label{fig:vmin_eta_mchi_5}}
  \end{center}
\end{figure}

\begin{figure}
\begin{center}
  \includegraphics[width=0.8\textwidth]{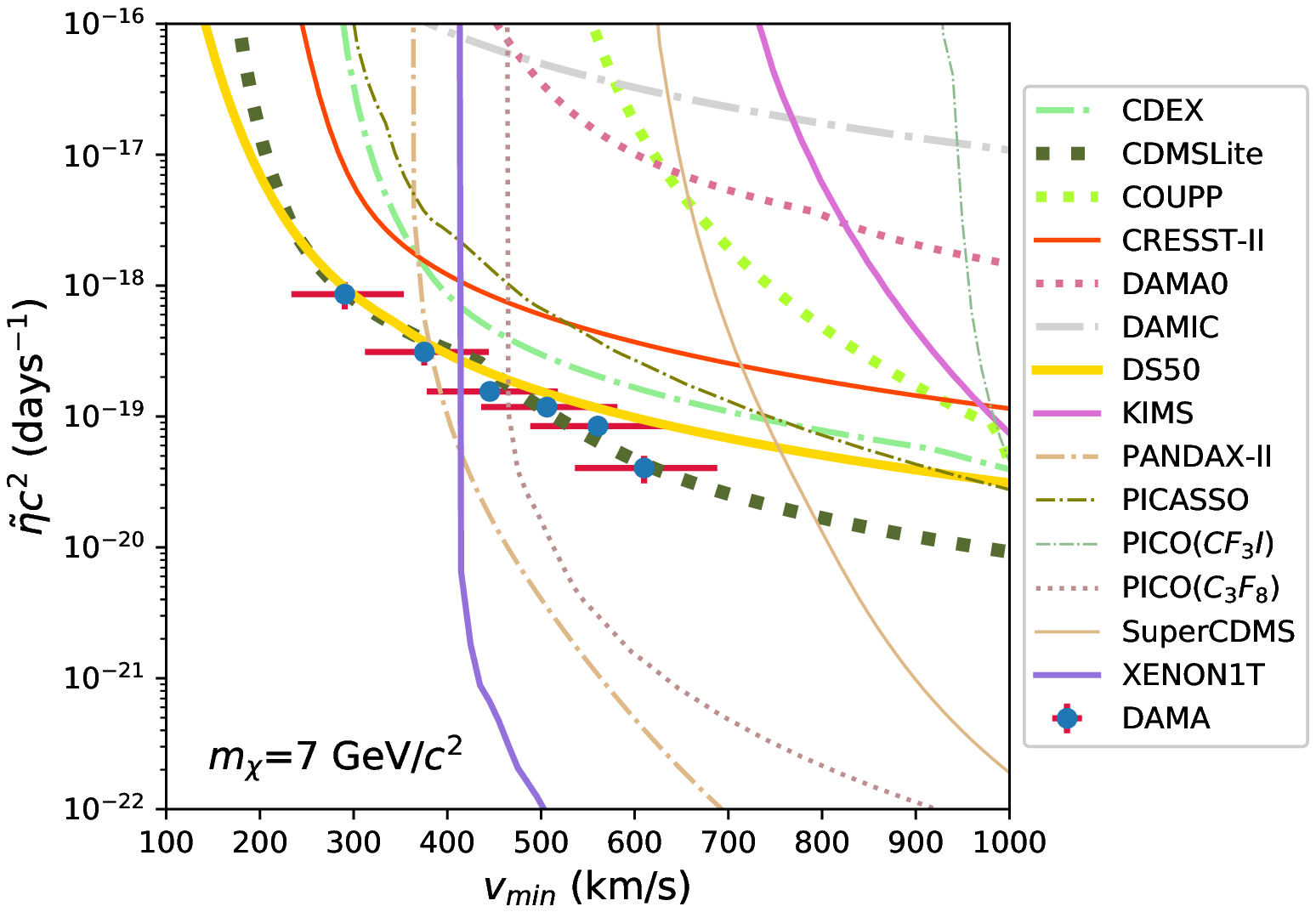}
  \caption{Same as Fig.\ref{fig:vmin_eta_mchi_5} but for
    $\mDM = 7~\GeV/c^2$.\label{fig:vmin_eta_mchi_7}}
  \end{center}
\end{figure}

\begin{figure}
\begin{center}
  \includegraphics[width=0.8\textwidth]{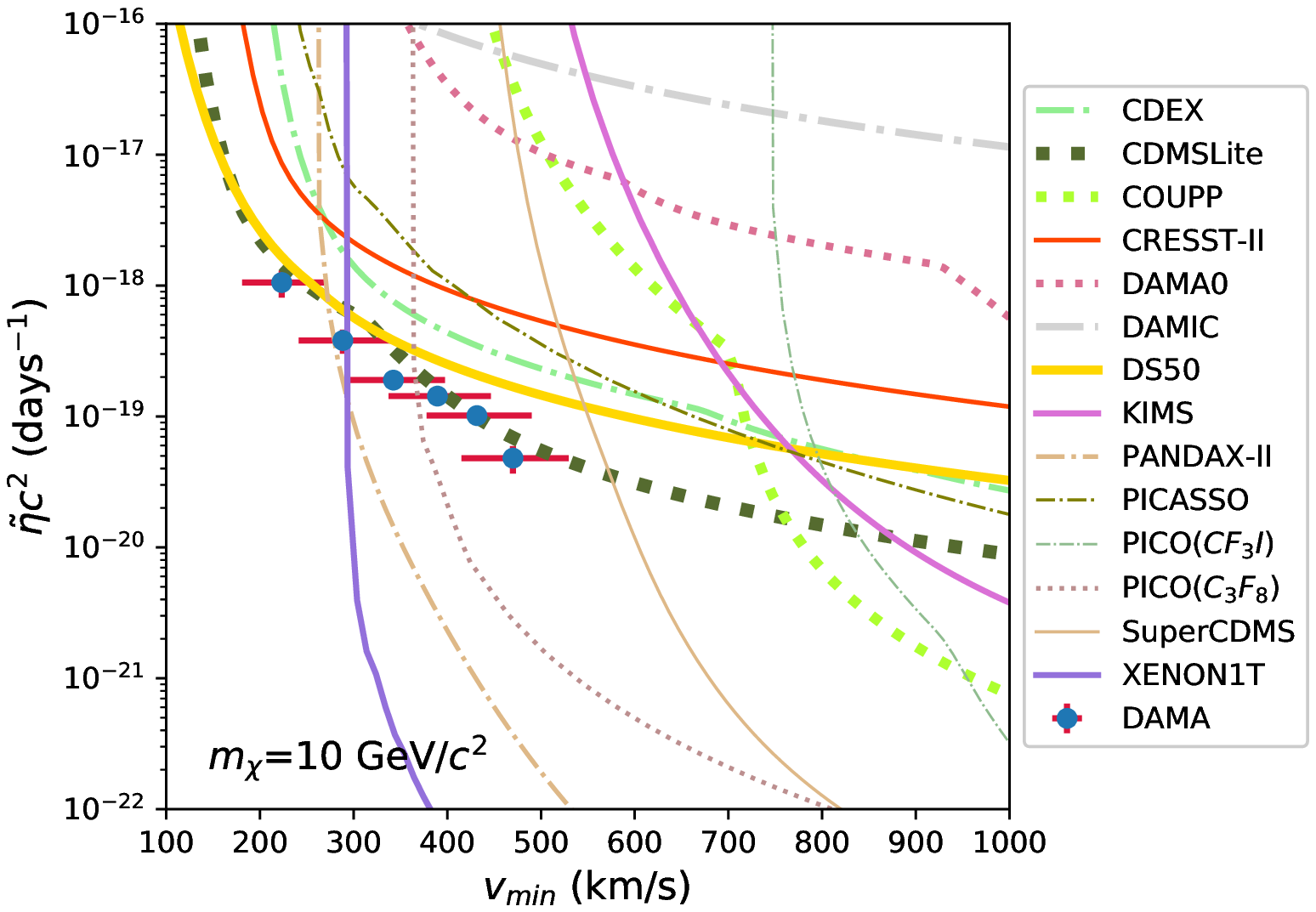}
  \caption{Same as Fig.\ref{fig:vmin_eta_mchi_5} but for
    $\mDM = 10~\GeV/c^2$.\label{fig:vmin_eta_mchi_10}}.
  \end{center}
\end{figure}

For the DAMA modulation, estimates of $\tilde\eta^1$ as the
regularized average $ \vmin^{-r} \langle \vmin^{r}
\tilde{\eta}^1\rangle_{[v_{{\rm min},1},v_{{\rm min},2}]} $ with
$r=10$ are shown as crosses in Figs.~\ref{fig:vmin_eta_mchi_5},
\ref{fig:vmin_eta_mchi_7} and~\ref{fig:vmin_eta_mchi_10}, where
$m_{\chi}=5$, 7 and $10~\GeV/c^2$, respectively. In these figures we
show the DAMA modulation amplitude in the first 6 energy bins of
\cite{dama_2018} from 1 keVee to 6 keVee, where the modulation signal
is concentrated. To determine the $\vmin$ interval corresponding to
each detected energy interval $[\Ed_1, \Ed_2]$ in DAMA we choose to
use $68\%$ central quantile intervals of the modified response
function $w^{(r)}_{[\Ed_1, \Ed_2]}(\vmin)$, i.e,, we determine
${\vmin}_{,1}$ and ${\vmin}_{,2}$ such that the areas under the
function $\vmin^{-r} \eR_{[\Ed_1, \Ed_2]}(\vmin)$ to the left of
${\vmin}_{,1}$ and to the right of ${\vmin}_{,2}$ are each separately
$16\%$ of the total area under the function. This gives the horizontal
width of the crosses corresponding to the rate measurements in
Figs.~\ref{fig:vmin_eta_mchi_5}, \ref{fig:vmin_eta_mchi_7} and
\ref{fig:vmin_eta_mchi_10}. On the other hand, the horizontal
placement of the vertical bar in the crosses corresponds to the
average of $\vmin$ using weights $w^{(r)}_{[\Ed_1, \Ed_2]}(\vmin)$,
i.e., $\vmin(\text{vertical bar})=\big[\int_0^\infty \ud\vmin
  \,\vmin^{1-r} \eR_{[\Ed_1, \Ed_2]}(\vmin)\big]/\big[\int_0^\infty
  \ud\vmin \, \vmin^{-r} \eR_{[\Ed_1, \Ed_2]}(\vmin)\big]$. The
extension of the vertical bar shows the $1 \sigma$ interval around the
central value of the measured rate.

To compute upper bounds on $\tilde{\eta}^{\,0}$ from
upper limits $R_{[\Ed_1, \Ed_2]}^{\rm lim}$ on the unmodulated rates,
we follow the conservative procedure in
Ref.~\cite{factorization}. Since $\tilde{\eta}^{\,0}(\vmin)$ is by
definition a non-decreasing function, the lowest possible
$\tilde{\eta}^0(\vmin)$ function passing through a point
$(v_0,\tilde{\eta}^{\,0})$ in $\vmin$ space is the downward step
function $\tilde{\eta}^0 \, \theta(v_0 - \vmin)$. The maximum value of
$\tilde{\eta}^0$ allowed by a null experiment at a certain confidence
level, denoted by $\tilde{\eta}^{\rm lim}(v_0)$, is then determined by
the experimental limit on the rate $R_{[\Ed_1, \Ed_2]}^{\rm lim}$ as
\begin{equation}
\tilde{\eta}^{\rm lim}(v_0) = \frac{R^{\rm lim}_{[\Ed_1,
      \Ed_2]}}{\int_0^{v_0} \ud \vmin \, \eR_{[\Ed_1, \Ed_2]}(\vmin)}\ .
\label{eq:etalim}
\end{equation}
These upper limits are shown as continuous lines in Figs.~\ref{fig:vmin_eta_mchi_5}, \ref{fig:vmin_eta_mchi_7} and
\ref{fig:vmin_eta_mchi_10} for the experiments listed there and in Appendix A.

We see that the DAMA points lie either in the excluded region or just
at its boundary (determined by the constraints from DS50, CDMSLite,
PANDAX--II and XENON1T). The best we could find in terms of
compatibility between DAMA and the other experiments are the two
lowest energy DAMA bins barely outside the excluded region at $m_\chi
\approx 7$--$8~\GeV/c^2$. In particular, it appears impossible to
explain all modulated bins in DAMA with anapole dark matter and at the
same time account for the other null direct DM search results, even in
the context of a halo-independent analysis. This is in sharp contrast
to the situation four years ago when anapole dark matter was still
viable when analyzed in a halo--independent way~\cite{anapole_2014}.

To be pedantic, one could object that we are actually comparing two
quantities defined differently for the modulated and unmodulated parts
of the DM signal, namely Eq.~(\ref{eq:eta_crosses}) and
Eq.~(\ref{eq:etalim}). For the two lowest DAMA energy bins that lie
near the boundary of the excluded region in
Figs.~\ref{fig:vmin_eta_mchi_5}--\ref{fig:vmin_eta_mchi_10} (near the
CDMSlite, DS50, and PANDAX-II upper limits), one may want to consider
more sophisticated analysis methods in which such objection is avoided
(e.g., the method of Ref.~\cite{Gondolo:2017jro}). On the other hand,
the DAMA bins above 2 keVee are excluded by XENON1T, PANDAX-II, and
PICO(C${}_3$F$_{8}$) by several orders of magnitude, and a more
sophisticated analysis is not warranted.

\section{Conclusions}
\label{sec:conclusions}

We have re--examined the case of anapole dark matter as an explanation for the DAMA annual modulation in light of the DAMA/LIBRA--phase2 results and improved upper limits from other DM searches.

For a Maxwellian WIMP velocity distribution, anapole dark matter is unable to provide an explanation of the DAMA modulation compatible with the other direct DM search results. Nevertheless, anapole dark matter
provides a better fit to the DAMA--phase2 modulation data than a
a standard isoscalar spin--independent interaction. This is due to the contribution from the magnetic moment of sodium, which reduces the hierarchy between the ADM
WIMP--iodine and WIMP--sodium cross sections compared to the SI case.

A halo-independent analysis shows that explaining the DAMA modulation
above 2 keVee in terms of anapole dark matter is basically impossible
in the face of the null results of XENON1T, PANDAX-II, and
PICO(C${}_3$F$_{8}$). On the other hand, the DAMA/LIBRA--phase2
modulation measurements below 2 keVee lie near the border of the
excluded region.

We conclude that in light of current measurements, anapole dark matter
does not seem to be a viable explanation for the totality of the DAMA
modulation, not even in a halo--independent analysis, although the
DAMA/LIBRA--phase2 modulation measurements below 2 keVee are
marginally allowed.

\acknowledgments This research was supported by the Basic Science
Research Program through the National Research Foundation of Korea
(NRF) funded by the Ministry of Education, grant number
2016R1D1A1A09917964. The work of P.G.\ was partially supported by NSF
Award PHY-1720282. P.G. thanks Sogang University for the kind and
gracious hospitality during the course of this work.

\appendix
\section{Experiments}
  \label{app:exp}

  In our analysis we have included an extensive list of updated
  constraints from existing DM direct-search experiments:
  CDEX~\cite{cdex},
  CDMSlite~\cite{cdmslite_2017},
  COUPP~\cite{coupp},
  CRESST-II~\cite{cresst_II,cresst_II_ancillary},
  DAMIC~\cite{damic},
  DAMA (modulation data~\cite{dama_1998, dama_2008,dama_2010,dama_2018} and
  average count rate~\cite{damaz}, indicated as DAMA0 in the plots),
  DarkSide--50~\cite{ds50} (indicated as DS50 in the plots),
  KIMS~\cite{kims_2014},
  PANDAX-II~\cite{panda_2017},
  PICASSO~\cite{picasso},
  PICO-60 (using a CF${}_3$I target~\cite{pico60_2015} and a
  C${}_3$F${}_8$ target~\cite{pico60}),
  SuperCDMS~\cite{super_cdms_2017} and
  XENON1T~\cite{xenon_1t}.
  With the exception of the latest result from
  XENON1T~\cite{xenon_2018}, the details of the treatment of the other
  constraints are provided in the Appendix of
  ~\cite{sensitivities_2018}.  For XENON1T (2018 analysis), we have
  assumed 7 WIMP candidate events in the range of 3PE $ \le S_1 \le $
  70PE, as shown in Fig.~3 of Ref.~\cite{xenon_2018} for the primary
  scintillation signal S1 (directly in Photo Electrons, PE), with an
  exposure of 278.8 days and a fiducial volume of 1.3 ton of xenon. We
  have used the efficiency taken from Fig.~1 of~\cite{xenon_2018} and
  employed a light collection efficiency $g_1$=0.055; for the light
  yield $L_y$ we have extracted the best estimation curve for photon
  yields $\langle n_{ph} \rangle /E$ from Fig.~7
  in~\cite{xenon_2018_quenching} with an electric field of $90~{\rm
    V/cm}$.

\end{document}